\documentclass[onecolumn,showpacs,preprintnumbers,amsmath,amssymb,aps]{revtex4}
\def \be{\begin{equation}}
\def \ee{\end{equation}}
\def \bea{\begin{eqnarray}}
\def \eea{\end{eqnarray}}
\def \ban{\begin{align}}
\def \ean{\end{align}}
\def \ben{\begin{enumerate}}
\def \een{\end{enumerate}}
\def \bit{\begin{itemize}}
\def \eit{\end{itemize}}

\def \unit{\leavevmode\hbox{\small1\kern-3.6pt\normalsize1}}
\def\lsim{\raise0.3ex\hbox{$\;<$\kern-0.75em\raise-1.1ex\hbox{$\sim\;$}}}
\def\gsim{\raise0.3ex\hbox{$\;>$\kern-0.75em\raise-1.1ex\hbox{$\sim\;$}}}
 \normalsize
\newcommand{\Rp}{\mbox{$\not \hspace{-0.15cm} R_p$}}

\newcommand{\bmat}{\left(\begin{array}}
\newcommand{\emat}{\end{array}\right)}

\usepackage{graphicx}
\usepackage{dcolumn}
\usepackage{bm}

\begin{document}

\preprint{}

\title{SUSY $R$ parity violation and CP asymmetry in semi-leptonic $\tau$-decays \\}

\author{D. Delepine}
 \email{delepine@fisica.ugto.mx}
\affiliation{Instituto de F\'isica de la Universidad de Guanajuato,
 C.P. 37150, Le\'on, Guanajuato, M\'exico.}%

\author{G. Faisel}
 \email{gaiber_faisel@bue.edu.eg}
\author{S. Khalil}%
\email{skhalil@bue.edu.eg} \affiliation{Ain Shams University,
Faculty of Science, Cairo 11566, Egypt.} \affiliation{Center for
Theoretical Physics, British University in Egypt, Shorouk city,
Cairo, 11837, Egypt.}

\date{\today}


\begin{abstract}
We analyze the CP violation in the semileptonic $\vert \Delta S
\vert=1$ $\tau$-decays in supersymmetric extensions of the
standard model (SM) with $R$  parity violating term.
 We show that the CP asymmetry of $\tau$-decay is
enhanced significantly and the current experimental limits
obtained by CLEO collaborations can be easily accommodated. We
argue that observing CP violation in semi leptonic $\tau$-decay would be a
clear evidence for $R$-parity violating SUSY extension of the SM

\end{abstract}


\maketitle
\section{Introduction}

CP violation is one of the main open questions in high energy
physics. In the standard model (SM), all CP violating observables
should be explained by one complex phase $\delta_{CKM}$ in the
quark mixing matrix. The effect of this phase has been first
observed in kaon system and recently confirmed in $B$ decays. The
quark-lepton symmetry suggests that the lepton mixing matrix
should also violate CP invariance. However, the situation in the
lepton sector is very different. The only evidence for flavor
violation in this sector comes from the neutrino oscillations and
there is no, so far, any confirmation for CP violation in leptonic
decays. Hence, measuring of $CP$ asymmetry in $\tau$ decays will
open new window to study the $CP$ violation.

Within the SM, the direct $CP$ asymmetry rate of $ \tau^{\pm} \to
K^{\pm}\pi^0\nu$ is of order $O(10^{-12})$ \cite{Delepine:2005tw}.
This nearly vanishing asymmetry implies that the observation of
$\tau$-decay would be a clear signal for the presence of CP
violation beyond the SM.

Supersymmetry is one of the most interesting candidates for
physics beyond the SM. Supersymmetry provides a new sources of CP
violation through complex couplings in the soft SUSY breaking
terms. The CP asymmetry in the decay mode $ \tau^{\pm} \to
K^{\pm}\pi^0\nu_{\tau}$, in minimal supersymmetric extension of
the SM with $R$ parity conservation, has been computed in Ref.
\cite{gaiber}. It was shown that the SUSY contributions can
enhance the CP asymmetry rate in $\tau$-decay by many order of
magnitude. However, the typical value is of order $O(10^{-7})$
which is still much smaller than the current experimental bound.

The aim of this paper is to show that a significant enhancement
for the CP asymmetry of $ \tau^{\pm} \to K^{\pm}\pi^0\nu_{\tau}$
can be obtained in SUSY models with $R$ parity violating terms. It
turns out that the $R$ parity violating terms (in particular, the
Lepton number violating ones) induce a tree level contribution to
$\tau$-decay. This contribution is proportional to the $R$ parity
couplings $\lambda$ and $\lambda'$, which in general are complex.
We find that this new source of CP violation enhance the asymmetry
of $\tau$ decay. We impose new constraints on the couplings
$\lambda$ and $\lambda'$ from the experimental limits, obtained by
CLEO collaborations\cite{Kuhn:1996dv,pipi,cleo,Bonvicini:2001xz}.

 The paper is organized as follows. In section 2, some general features on $\tau$ semi-leptonic decays are recalled.
 Section 3 is devoted for analyzing the CP asymmetry of
 $\tau^{\pm} \to K^{\pm}\pi^0\nu$ in SUSY models with $R$ parity
 violation. We derive the corresponding effective hamiltonian
 and show that terms that violate $R$ parity may give significant contribution to the CP asymmetry in
 $\tau$-decay. Finally, we give our conclusions in section 4.

\section{CP asymmetry of $\tau$ semi-leptonic decay in MSSM. }

 In this section we analyze the $CP$ asymmetry of $\tau$ semileptonic
 decay modes within MSSM, we will focus on the decay mode $ \tau^{\pm} \to K^{\pm}\pi^0\nu$.
 The general amplitude for $\tau^-(p) \rightarrow
K^-(k)\pi^0(k')\nu_{\tau}(p')$ is given by
 \begin{eqnarray}
  {\cal M} &=&
\frac{G_FV_{us}}{\sqrt{2}}
\left\{\bar{u}(p')\gamma^{\mu}(1-\gamma_5)u(p) F_V(t)
\left[(k-k')_{\mu}-\frac{\Delta^2}{t}q_{\mu} \right] \right. \nonumber \\
& & + \bar{u}(p')(1+\gamma_5)u(p) m_{\tau} \Lambda
F_S(t)\frac{\Delta^2}{t} \nonumber
\\
& & \left. + F_{T}\langle K\pi |\bar{s}\sigma _{\mu \upsilon
}u|0\rangle \bar{u}(p^{\prime })\sigma ^{\mu \upsilon
}(1+\gamma_5) u(p)\right\}, \nonumber \end{eqnarray}%
where $q=k+k'$ ($t=q^2$) is the momentum transfer to the hadronic
system, $\Delta^2   \equiv m_K^2-m_{\pi}^2$ and $F_{V,S,T}(t)$ are
the {\it effective} form factors describing the hadronic matrix
elements. In SM, $F_T=0$, $\Lambda=1$

\begin{eqnarray}
\sum_{pols}\vert\mathcal{M}\vert^2&\sim &\vert
F_V\vert^2(2p.Qp^{\prime}.Q-p.p^{\prime}Q^2)+\vert\Lambda \vert^2
\vert
F_S(t)\vert^2M^2p.p^{\prime}  \nonumber \\
&& +2Re\Lambda\cdot Re(F_SF_V^*)M m_{\tau}p^{\prime}.Q
-2Im\Lambda\cdot Im(F_SF_V^*)Mm_{\tau}p^{\prime}.Q ~,\nonumber
\end{eqnarray}
where $Q_{\mu}=(k-k')_{\mu}-\frac{\Delta^2}{t}q_{\mu}$.

The last two terms disappear once we integrate on the kinematical
variable $u$ unless the form factor have a $u$ dependence. The form factors $F_{V,S}(t)$ can receive weak phase
through higher order contributions and
hence it is  possible to generate a CP asymmetry in total decay
rates but within SM this CP asymmetry is nearly vanishing\cite{Delepine:2005tw}.

The CP asymmetry is defined as

\begin{eqnarray}
A_{CP}&=&\frac{\Gamma(\tau^+ \to
K^+\pi^0\bar{\nu}_{\tau})-\Gamma(\tau^- \to
K^-\pi^0\nu_{\tau})}{\Gamma(\tau^+ \to
 K^+\pi^0\bar{\nu}_{\tau})+\Gamma(\tau^-
\to K^-\pi^0\nu_{\tau})}. \nonumber
\end{eqnarray}

The effective Hamiltonian $H_{eff}$ derived from SUSY
superpotential with $R$ parity symmetry conserved can be expressed
as \cite{gaiber}
\begin{equation}
H_{eff}=\frac{G_{F}}{\sqrt{2}}V_{us}\sum_{i}C_{i}(\mu )Q_{i}(\mu
), \label{SHeff}
\end{equation}%
where $C_{i}$ are the Wilson coefficients and $Q_{i}$ are the
relevant local operators at low energy scale $\mu \simeq m_{\tau
}$. The operators are given by
\begin{eqnarray}
Q_{1} &=&(\bar{\nu}\gamma ^{\mu }L\tau )(\bar{s}\gamma _{\mu }Lu), \\
Q_{2} &=&(\bar{\nu}\gamma _{\mu }L\tau )(\bar{s}\gamma _{\mu }Ru), \\
Q_{3} &=&(\bar{\nu}R\tau )(\bar{s}Lu), \\
Q_{4} &=&(\bar{\nu}R\tau )(\bar{s}Ru), \\
Q_{5} &=&(\bar{\nu}\sigma _{\mu \nu }R\tau )(\bar{s}\sigma ^{\mu
\nu }Ru).
\end{eqnarray}%
where $L,R$ are defined as $L,R =1\mp \gamma_5$ and $\sigma ^{\mu \nu }=%
\frac{i}{2}[\gamma ^{\mu },\gamma ^{\nu }]$. The SUSY contributions to the Wilson coefficients $C_{i}$%
can be found in Ref.\cite{gaiber}. For dominant $C_{3}$ and/or
$C_{4}$, one finds that the decay amplitude is given by%
\begin{eqnarray}
\mathcal{A}_{T}(\tau \rightarrow K\pi \nu ) &=&\frac{G_{F}V_{us}}{\sqrt{%
2}}(1+C_{1})\times   \nonumber \\
&&\!\!\!\!\!\!\!\!\left\{ f_{V}Q^{\mu }\bar{u}(p^{\prime })\gamma
_{\mu
}Lu(p)+\left[ m_{\tau }+\left( \frac{C_{3}+C_{4}}{1+C_{1}}\right) \frac{t}{%
m_{s}-m_{u}}\right] f_{S}\bar{u}(p^{\prime })Ru(p)\right\} \
.\nonumber
\end{eqnarray}

Using CLEO limit, we can translate this bound into:
\begin{equation} -0.010\leq Im\left(
\frac{C_{3}+C_{4}}{1+C_{1}}\right) \leq 0.004\ ,\nonumber
\end{equation}%
where we have used $m_{s}-m_{u}=100$ MeV, and the average value
$\langle t\rangle \approx (1332.8\ \mbox{\rm MeV})^{2}$. However,
for $M_{1}=100$ and $M_{2}=200$ GeV \ and $\mu =M_{\tilde{q}}=400$
GeV and $\tan \beta =20$, one gets
\begin{equation} Im\left(
\frac{C_{3}+C_{4}}{1+C_{1}}\right) \simeq 1.3\times
10^{-5}Im(\delta _{21}^{d})_{RL} \nonumber
\end{equation}
It is worth noting that the mass insertions $(\delta_{21}^d)_{RL}$ are constrained by the $%
\Delta M_K $ and $\epsilon_K $ as follows:
\begin{eqnarray}
\vert(\delta^d_{21})_{RL}\vert \raise0.3ex\hbox{$\;<$\kern-0.75em%
\raise-1.1ex\hbox{$\sim\;$}} 4 \times 10^{-3}.
\end{eqnarray}
Therefore, the resultant CP asymmetry of $\tau \rightarrow K\pi
\nu$ is smaller, by few order of magnitude, than the current
experimental limit.

%
\section{$\tau$ decay $CP$ asymmetry in SUSY without $R$ parity}
In this section we study the effect of including terms that
violate lepton and baryon number on the tau decay $CP$ asymmetry.
The gauge invariance does not insure the conservations of both
baryon number and lepton number and hence we can allow the SUSY
superpotential to have the $R$ parity violating terms. on the
other hand side, $R$ parity violation can be motivated by some
controversial experimental observations, like events with missing
energy and a hadron jet in the $H1$ experiment at HERA\cite{Choi:2006ms}. Recall
that the most general superpotential that violates the $R$ parity
symmetry can be written as \cite{Barger:1989rk} %
\bea W_{\Rp}\ &=&\frac{1}{2} \lambda_{ijk} L_iL_j\bar{E_k} +
\lambda^{'}_{ijk} L_iQ_j\bar{D_k} + \frac{1}{2} \lambda^{''}_{ijk}
\bar{U}_i\bar{D}_j\bar{D}_k+\kappa_iL_iH_2, %
\eea %
where a summation over the generation indices $i,j=1,2,3$ and over
gauge indices is understood.  The $\lambda_{ijk}$ is
anti-symmetric in $\{ i,j\}$ because of the contraction of $SU(2)$
indices and hence $\lambda_{ijk}$ are non-vanishing only for
$i<j$. The $\lambda''_{ijk}$ is anti-symmetric in $\{j,k\}$.
Therefore $j\not= k$ in $\bar{U_i}\bar{D_j}\bar{D_k}$ and hence we
can write the superpotential $W_{\Rp}$ as:%
\bea %
W_{\Rp}\ &=& \lambda_{ijk} L_iL_j\bar{E_k} + \lambda^{'}_{ijk}
L_iQ_j\bar{D_k} + \lambda^{''}_{ijk}
\bar{U}_i\bar{D}_j\bar{D}_k+\kappa_iL_iH_2. %
\eea%

\begin{figure}
\includegraphics[width=4cm]{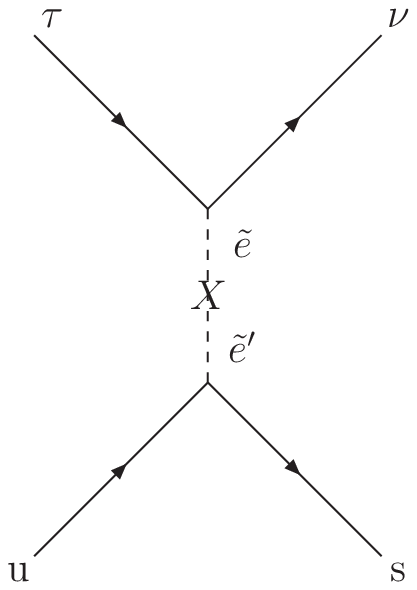}~~~~~~~~~~~~~\includegraphics[width=4cm]{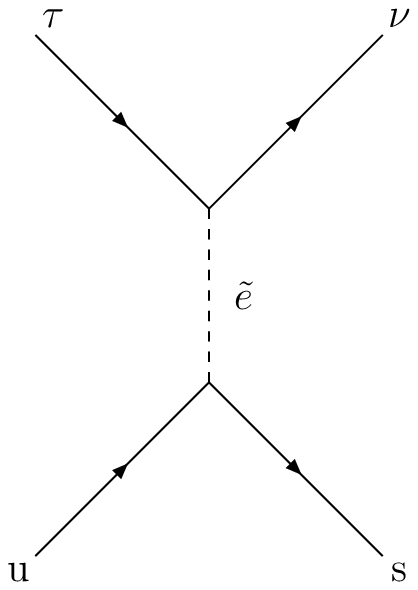}\\
\caption{$R$ parity violation SUSY contributions to $\protect\tau ^{-}\rightarrow \bar{u}s%
\protect\nu _{\protect\tau }$ transition.} %
\label{Rdiagrams}
\end{figure}

To insure that the proton is stable, we require only the
conservation of baryon number and hence we forbid the term
$\lambda''_{ijk} \bar{U_i}\bar{D_j}\bar{D_k}$.
 Expanding  $W_{\Rp}\ $  term into the Yukawa couplings yields %
\bea%
{\cal L} &=&\lambda_{ijk}\left[ {\tilde\nu}^i_L{\bar e}^k_Re_L^j
+{\tilde e}_L^j{\bar e}_R^k\nu^i_L + ({\tilde e}_R^k)^*
({\bar\nu}_L^i)^ce_L^j - (i\leftrightarrow j)\right]\nonumber
\\&+&\lambda'_{ijk}\left[ {\tilde\nu}^i_L{\bar d}^k_Rd_L^j +{\tilde
d}_L^j{\bar d}_R^k\nu^i_L + ({\tilde d}_R^k)^*
({\bar\nu}_L^i)^cd_L^j - {\tilde e}^i_L{\bar d}^k_Ru_L^j-{\tilde
u}^i_L{\bar d}^k_Re_L^j\right. \nonumber\\
&-& \left.({\tilde d}_R^k)^*({\bar e
}_L^i)^cu_L^j\right] + h.c.,~~ %
\eea%
where, tilde denotes the scalar fermion superpartners. The leading
diagrams for $\tau \to k \pi \nu$ are illustrated in
Fig.\ref{Rdiagrams}.

The corresponding effective Hamiltonian, $H_{eff}$, derived from
SUSY $R$ parity violating terms can be expressed as \be H_{eff}=
\frac{G_F}{\sqrt{2}}V_{us}\sum_i C_i(\mu) Q_i(\mu),\ee where $C_i$
are the Wilson coefficients and $Q_i$ are the relevant local
operators at low energy scale $\mu\simeq m_\tau$. These operators
are given by %
\bea %
Q_1 &=&(\bar{\nu} \gamma^{\mu} L \tau)(\bar{s}\gamma_{\mu} L u),\\
Q_2&=&(\bar{\nu} R \tau)(\bar{s} L u),\\  Q_3 &=&(\bar{\nu} L
\tau)(\bar{s} L u).%
\eea

The Wilson coefficients $C_i$, at the electroweak scale, can be
expressed as $C_i = C_i^{SM} + C_i^{SUSY}$. For $i=2,3$ $C^{SM}_i$
vanish identically. In this respect, the Wilson coefficients $C_i$
are given by%
\bea %
C_1 &=& 1\\
C_2&=&\frac{\sqrt{2}}{G_F
V_{us}}\left(\frac{1}{\tilde{m}^2}\right)(\lambda_{133}\lambda'_{312}
+\lambda_{123}\lambda'_{212}),\\
C_3 &=& \frac{\sqrt{2}}{G_F
V_{us}}\left(\frac{1}{\tilde{m}^4}\right)(\lambda_{13k}\lambda'_{i12}
)(\delta_{LR}^l)_{ik}~.%
\eea%

 The couplings $\lambda_{ijk}$ and $\lambda'_{ijk}$ are generally complex numbers.
 At a value for $\tilde{m}=100 GeV$, the upper bounds on these couplings are given
 as follows \cite{Dreiner:1998wm}: %
 \bea%
 |\lambda_{131}|&=&|\lambda_{132}|=0.06 , \nonumber\\
 |\lambda_{133}|&=& 0.004, ~~~~~~ |\lambda_{123}|= 0.05,\nonumber\\
 |\lambda'_{212}|&=& 0.09,~~~~~~~~ |\lambda'_{312}|=0.16,\label{upperbounds}\\
 |\lambda'_{112}|&=& 0.02,~~~~~~~~ |\lambda'_{312}|=0.16.\nonumber%
 \eea
 As can be seen, $C_3$ can be dropped comparing to $C_1$ and $C_2$.
 The decay amplitude for the decay $\tau^-(p) \to K^-(k)\pi^0(k')
\nu(p')$ including SM and SUSY contributions becomes:%
\be%
{\cal M} = \frac{G_F}{\sqrt{2}}V_{us}  \left\{ \langle K\pi
|\bar{s} \gamma_{\mu}u| 0 \rangle \bar{\nu}\gamma_{\mu}L \tau +
C_2\langle K\pi |\bar{s}u| 0 \rangle\ \bar{\nu}R \tau +C_3\langle
K\pi |\bar{s}u| 0 \rangle\ \bar{\nu}L \tau\right\}\,%
\ee %
Using %
\be%
\langle
K\pi|\bar{s}\gamma_{\mu}u|0\rangle=f_V(t)Q_{\mu}+f_S(t)(k+k')_{\mu}~,%
\ee %
and %
\be%
\langle K\pi|\bar{s}u|0\rangle=\frac{t}{m_s-m_u}f_S(t)~.\ee %
The amplitude can be written as: %
\bea %
{\cal A}_T(\tau \to K \pi
\nu)&=&\frac{G_FV_{us}}{\sqrt{2}}
\left\{f_V(t)Q^{\mu}\bar{\nu}(p')\gamma_{\mu}L\tau(p) \right.
\frac{}{} \nonumber\\ && \left. +\left[m_{\tau}+
  C_2 \frac{t}{m_s-m_u}\right]f_S(t)
\bar{\nu}(p')R \tau(p)\right\}.%
\eea

This expression should be compared with the decay amplitude given in
 Eq. (2) of Ref. \cite{Bonvicini:2001xz}:%
\be %
{\cal A}(\tau^-\to K \pi\nu_{\tau}) \sim
{\bar{u}(p')}\gamma_{\mu}Lu(p)f_V Q^{\mu}
 +\Lambda {\bar{u}(p')}Ru(p)f_SM\ ,
\ee%
where $M=1$ GeV is a normalization mass scale. Hence one finds
the relation: %
\be %
\Lambda M =m_{\tau}+ C_2\frac{t}{m_s-m_u}.%
\ee%
The first term in the last equation is the usual contribution of
the SM, which is real, and the second term arises from the SUSY
contributions.

Using the bound obtained by the CLEO collaboration : $-0.172 <
Im(\Lambda) < 0.067$ at  90\% C.L. \cite{cleo}, we can translate
this bound into: %
\be%
 -0.010<Im  C_2 < 0.004\ , %
\ee %
where we have used again as before, $m_s-m_u =100$ MeV, and
the average value $\langle t \rangle \approx (1332.8 \ \mbox{\rm
MeV})^2$. After substitution, we can write%
\be %
-0.010<Im \left[\frac{\sqrt{2}}{G_F
V_{us}}\left(\frac{1}{\tilde{m}^2}\right)\left(\lambda_{133}\lambda'_{312}
+\lambda_{123}\lambda'_{212}\right)\right] < 0.004\ .%
\ee%
For $G_F = 1.166\times 10^{-5}$ GeV, $\tilde{m}= 100$ GeV, and
$V_{us} = 0.22$, one obtains the following bound %
\be%
 -1.8\times10^{-4}<Im (\lambda_{133}\lambda'_{312}
+\lambda_{123}\lambda'_{212}) < 7.2\times10^{-5}\ . %
\ee %
From the upper bounds on the Yukawa couplings: $\lambda_{ijk}$ and
$\lambda'_{ijk}$, reported in Eq.(\ref{upperbounds}), one finds
that $ \vert \lambda_{133}\lambda'_{312}\vert \lsim 10^{-4}$,
while the coupling $\lambda_{123}$ is unconstrained. Thus, the
above equation leads to the following bound on $Im(\lambda_{123}
\lambda'_{212})$: %
\be%
-1.8\times10^{-4}<Im (\lambda_{123}\lambda'_{212}) <
7.2\times10^{-5}\ . %
\ee %
Thus, with order $10^{-2}$ complex Yukawa couplings
$\lambda_{123}$ and/or $\lambda'_{212}$, the CP asymmetry
experimental limits obtained by CLEO collaborations can be easily
accommodated. This result is an intrinsic feature for $R$ parity
SUSY contribution to the CP asymmetry of $\tau \to k \pi \nu$. It
is important to stress that this is the only model, to our
knowledge, that enhance CP asymmetry of $\tau$ decay significantly
and account for the CLEO limits. Therefore, a confirmation for
CLEO measurements would be a clear evidence of $R$ parity
violating SUSY extension of the SM.

\section{Conclusion}

We have studied the supersymmetric contributions to the CP
asymmetry of $\tau \to k \pi \nu$ decay. We emphasized that CP
asymmetry in this decay is nearly vanishing within the SM.
Therefore, any non-vanishing CP asymmetry in this decay channel
will be a clear evidence for physics beyond the SM. We have shown
how physics beyond standard model as supersymmetric extensions of
the SM could induce CP violating asymmetry in the double
differential distribution as CLEO collaboration did. In case of
Supersymmetry with conserved R parity, it has been found that the
CP asymmetry is enhanced by several orders of magnitude than the
SM expectations. However, the resulting asymmetry is still well
below the current experimental limits obtained by CLEO
collaborations\cite{gaiber}. Within R-parity violation SUSY
models, we  found that the CP asymmetry of $\tau$-decay is
enhanced significantly and the current experimental limits
obtained by CLEO collaborations can be easily accommodated.

\section{Acknowledgements}
We would like to thank Gabriel Lopez Castro for discussions. The
work of D.D. was partially supported by PROMEP UGTO-PTC project,
DINPO, Conacyt Project numero 46195. G. F. acknowledges support
from CIMO and hospitality of HIP and University of Helsinki, while
this work was in progress.



\end{document}